\documentclass[10pt]{myels}
\usepackage{epsfig}
\newcommand{\tetaot}{\mbox{$\theta_{13}$}}
\newcommand{\tetatt}{\mbox{$\theta_{23}$}}

\newcommand{\raw}{\rightarrow}

\newcommand{\be}{\begin{equation}}
\newcommand{\ee}{\end{equation}}
\newcommand{\bea}{\begin{eqnarray}}
\newcommand{\eea}{\end{eqnarray}}

\begin{document}

%
\begin{flushright}
{\tt hep-ph/0007281}\\
{UM-TH-00-19}
\end{flushright}
\vspace*{0.5cm}
%
%
%

\begin{frontmatter}
\begin{center}


\title{Summary of Golden Measurements at a $\nu$-Factory}



\begin{center}
\author[a]{A. \snm Cervera}
\author[b]{A. \snm Donini}
\author[b]{, M.B. \snm Gavela}
\author[a]{, J.J. \snm Gomez C\'adenas, }
\newline
\author[c]{P. \snm Hern\'andez}\footnote{On leave from Departamento de 
           F\'{\i}sica Te\'orica, Universidad de Valencia.} 
\author[b]{, O. \snm Mena}
\author[d]{ and S. \snm Rigolin}
\address[a]{Dep. de F\'{\i}sica At\'omica y Nuclear and IFIC, 
            Universidad de Valencia, \cny Spain}
\address[b]{Dep. de F\'{\i}sica Te\'orica, 
            Universidad Aut\'onoma de Madrid, 
            \cny Spain}
\address[c]{CERN, \cty 1211 Geneve 23, \cny Switzerland}
\address[d]{Dep. of Physics, University of Michigan, \cty Ann Arbor - 
            MI 48105, \cny USA}
\end{center}


%
\begin{abstract}
\noindent
The precision and discovery potential of a neutrino factory based 
on muon storage rings is summarized. For three-family neutrino oscillations, 
we analyze how to measure or severely constraint the angle $\theta_{13}$, 
CP violation, MSW effects and the sign of the atmospheric mass difference 
$\Delta m^2_{23}$. 
The appearance of ``wrong-sign muons'' at three reference baselines 
is considered: 732 km, 3500 km and 7332 km. We exploit the dependence of
the signal on the neutrino energy, and include as well realistic background 
estimations and detection efficiencies. The optimal baseline turns out to 
be $O$(3000 km). 
\end{abstract}
%
%
%
%
\end{center}
\end{frontmatter}
%
%
\pagestyle{plain} 
\setcounter{page}{1}
%
%
%
%
The atmospheric \cite{Superka,otheratm} plus solar \cite{sol} neutrino data 
point to neutrino oscillations and can be easily accommodated in a 
three-family mixing scenario. 
The new SuperK data seem to indicate a slight preference for the solar 
LMA-MSW solution and disfavor oscillations into sterile neutrinos both in 
the solar and atmospheric sectors \cite{newSK}. 
These facts improve the prospectives for a neutrino factory. 
In ten years from now, the planned experiments will possibly improve the 
precision in the solar and atmospheric sector and definitively exclude 
(or confirm) LSND signal. 
Nevertheless, there is a strong case for going further in the fundamental 
quest of the neutrino masses and mixing angles\footnote{For a four-family 
analysis at a neutrino factory see for example \cite{us99}.}. 
In fact no significant improvement is expected in the knowledge of: 
1) the sign of the atmospheric mass difference, $\Delta m_{23}^2$, 2)
the angle relating the solar and the atmospheric sectors, $\theta_{13}$, 
and 3) the CP-violating phase,~$\delta$. 

The most sensitive method to study these topics is to measure the transition 
probabilities involving $\nu_e$ and $\bar \nu_e$, in particular 
$\nu_e(\bar \nu_e) \raw \nu_\mu ( \bar \nu_\mu)$. 
This is precisely the golden measurement at the {\it neutrino factory} 
\cite{geer}. Such a facility is unique in providing high energy and intense 
$\nu_e (\bar \nu_e)$ beams coming from positive (negative) muons which decay 
in the straight sections of a muon storage ring \cite{firstmachine}. 
Since these beams contain also $\bar \nu_\mu (\nu_\mu)$ (but no 
$\nu_\mu (\bar \nu_\mu)$!), the transitions of interest can be measured by 
searching for ``wrong-sign'' muons \cite{dgh}. 

We will analyze in turn scenarios in which the solar oscillation lies in 
the SMA-MSW or VO range and in the LMA-MSW range \cite{golden,others}. In 
the latter, the dependence of the oscillation probabilities on the solar 
parameters $\theta_{12}, \Delta m^2_{12}$, and on the CP-odd phase, $\delta$,  
is sizeable at terrestrial distances and complicates the measurement of 
$\tetaot$ due to the presence of other unknowns (mainly $\delta$). 
The choice of the correct baseline is essential to disentangle $\tetaot$ 
and $\delta$. We shall consider the following 
``reference set-up'': neutrino beams resulting from the decay of 
$2 \times 10^{20} \mu^+$'s and/or $\mu^-$'s per year in a straight section 
of an $E_\mu = 50$ GeV muon accumulator. A long baseline (LBL) experiment 
with a 40 kT detector and five years of data taking for each polarity is 
considered. Alternatively, the same results could be obtained in one year 
of running for the higher intensity option of the machine, providing $10^{21}$ 
useful $\mu^+$'s and $\mu^-$'s per year. A realistic detector of magnetized 
iron \cite{cdg} will be considered and detailed estimates of the  
expected backgrounds and efficiencies included in the analysis. Three 
reference detector distances are discussed: 732 km, 3500 km and 7332 km. 
%
%
 
\begin{figure}
\begin{tabular}{lcr}
\hskip -0.5cm
\mbox{\epsfig{file=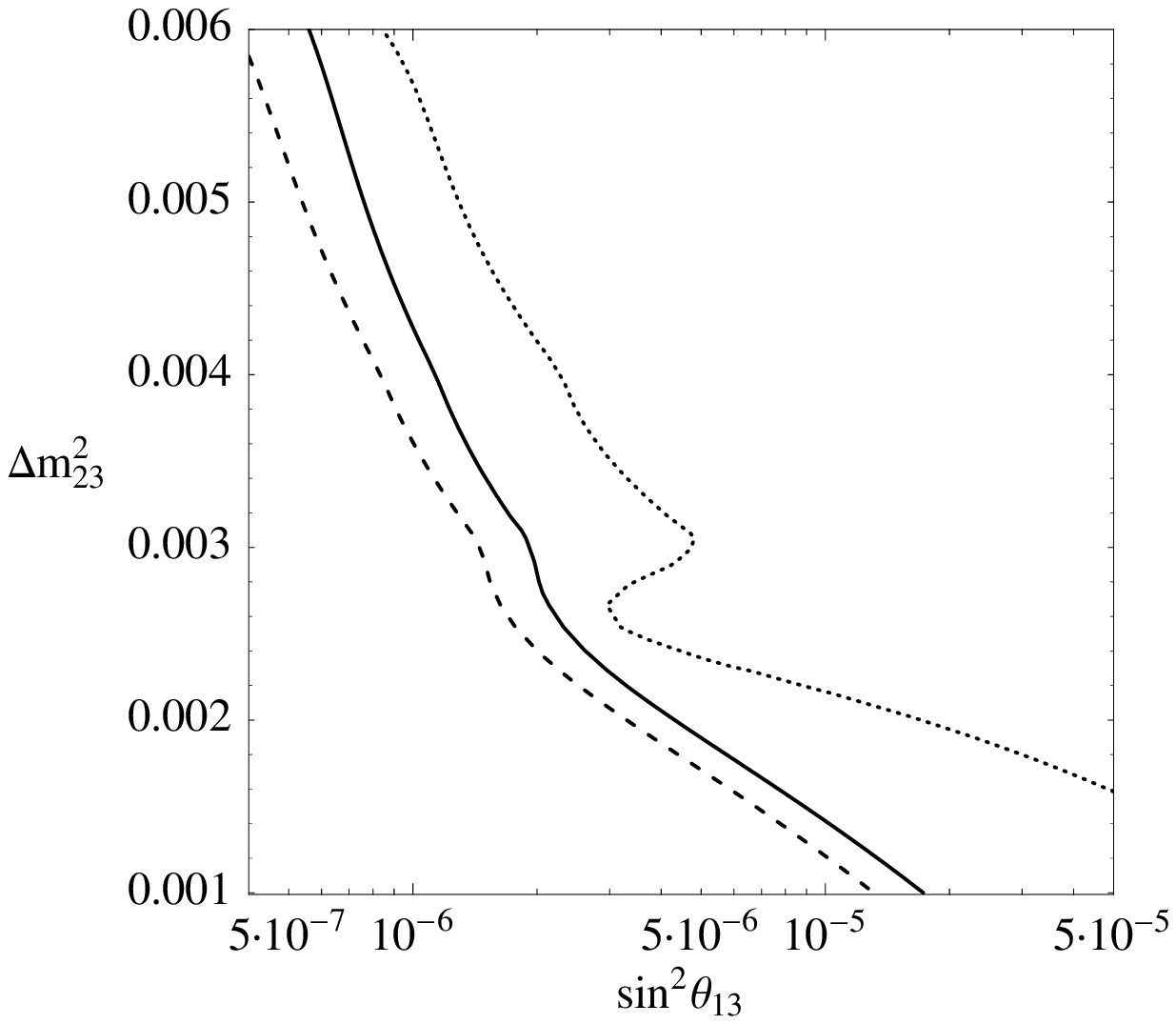,width=8cm} }
& \mbox{ \hskip -1cm} &
\mbox{\epsfig{file=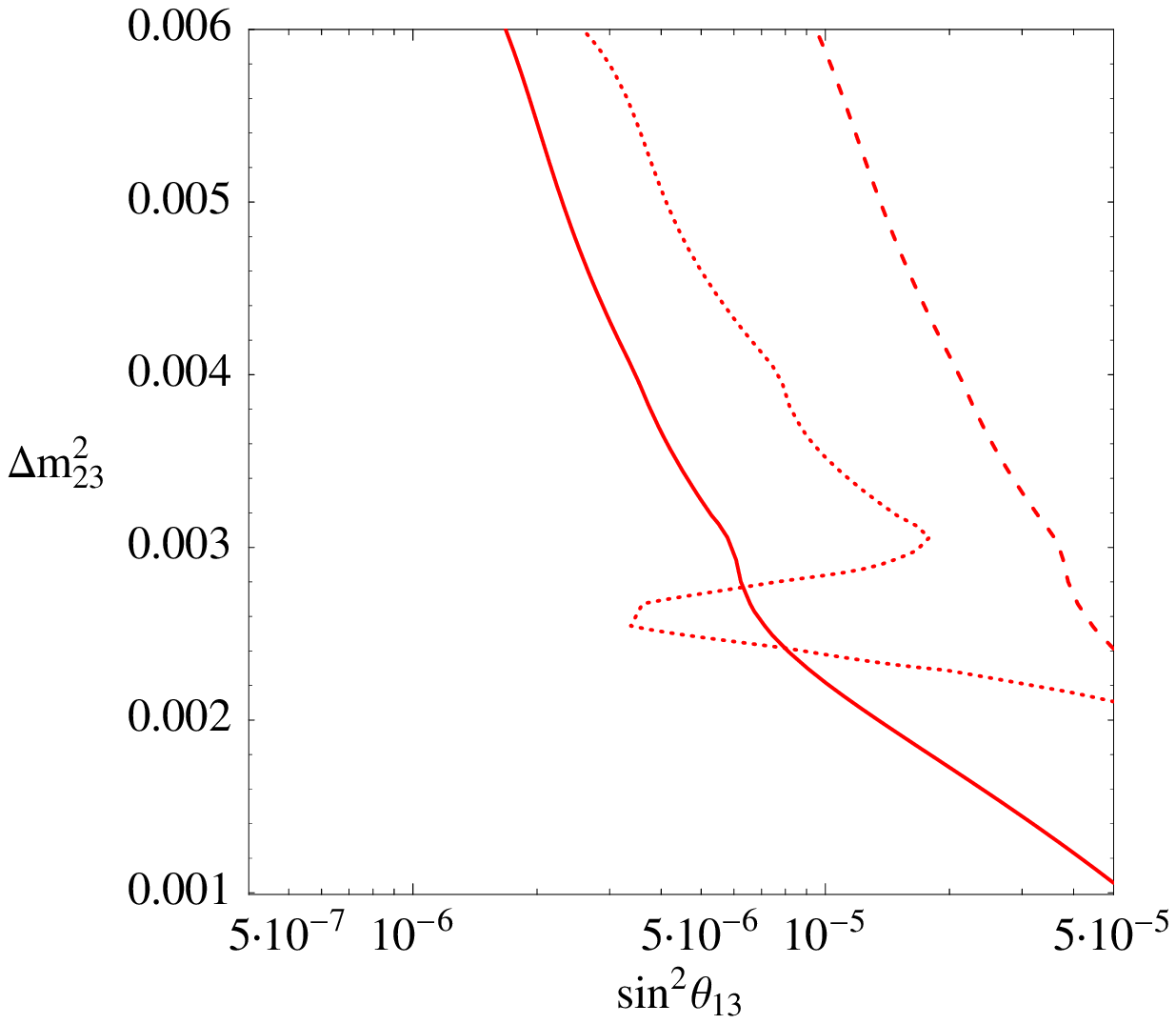,width=8cm} }
\end{tabular}
\caption{ \it Asymptotic sensitivity to $\sin^2 \tetaot$ as a function of 
$\Delta m^2_{23}$ at 90\% CL for $L=732$ km (dashed lines),
3500 km (solid lines) and 7332 km (dotted lines), in the SMA-MSW solution. 
Only statistical errors are included in the left plot while the right 
one include as well background errors and detection efficiencies.}  
\label{excluno}
\end{figure}

{\bf SMA-MSW or Vacuum solar deficit.} 
For the SMA-MSW or VO scenarios, the influence of solar parameters 
on the neutrino factory signals will be negligible\footnote{In practice, 
for the numerical results of this section, the central values in the 
SMA-MSW range are taken: $\Delta m_{12}^2 = 6 \times 10^{-6}$ eV$^2$ and 
$\sin^2 2 \theta_{12} = 0.006$.} and CP-violation out of reach. Besides 
its capability to reduce the errors on $\tetatt$ and $|\Delta m_{23}^2|$ 
to $\sim 1\%$ \cite{bgrw} the factory would still be a unique machine to 
constrain/measure $\tetaot$ \cite{dgh} and the sign of $\Delta m^2_{23}$. 
Consider first $\tetaot$. 
In Fig.~\ref{excluno} (left), we show the exclusion plot at 90\% CL, on the 
$\Delta m^2_{23}$ (in the range allowed by SuperK) versus $\sin^2 \tetaot$ 
plane, obtained with the full unbinned statistics and the two polarities. 
The same results, but including as well background errors and detection 
efficiencies are shown in Fig. \ref{excluno} (right). Notice that the 
sensitivity is better at $L =$ 3500 km than at 732 km when efficiencies 
and backgrounds are included. The latter are responsible for it.
The sensitivity at $7332$ km is also worse than at $3500$ km, due to the loss 
in statistics. In conclusion, the sensitivity to $\tetaot$ can be improved 
by three-four orders of magnitude with respect to the present limits. 

\begin{figure}
\begin{center}
\mbox{\epsfig{file=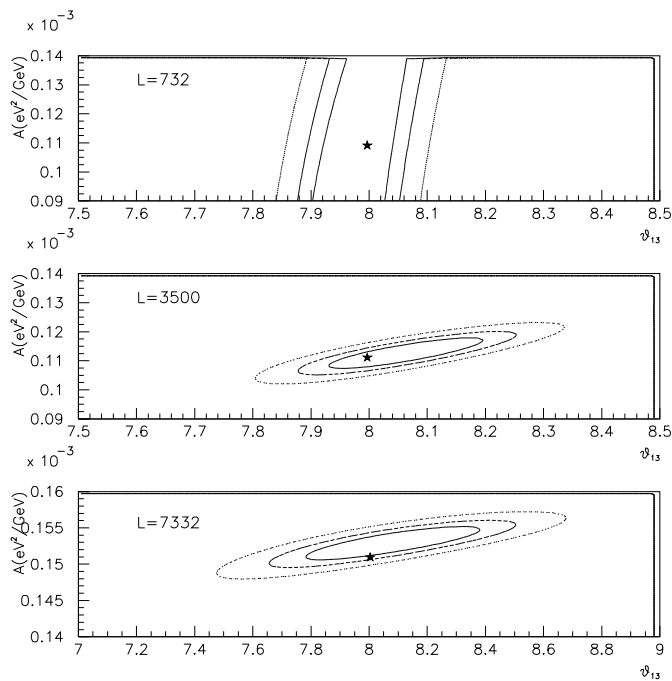,height=9cm}} 
\end{center}
\caption{\it 68.5, 90, 99 \%  CL contours resulting from a $\chi^2$ fit of  
$\tetaot$ and $A$. The parameters used to generate the ``data'' are denoted 
by a star, while the baseline(s) used in the fit is indicated in each plot. 
Statistical errors, backgrounds and efficiencies are included.}
\label{fig:sma8be}
\end{figure}
%

The second major topic would be of perform the first precise measurements 
related to matter effects, in order to determine the sign of $\Delta m^2_{23}$ 
We have studied the determination of the sign of $\Delta m^2_{23}$, 
assuming that the absolute value has by then been measured with a precision 
of 10\%. We have explored the region around the best fit values of SuperK: 
$|\Delta m_{23}^2|=2.8 \times 10^{-3}$ eV$^2$ and $\tetatt=45^\circ$.
We performed a $\chi^2$ analysis on the $\Delta m^2_{23}, \tetaot$ plane, 
as described in \cite{golden}. 
The conclusion is that, for ``data'' generated within the range 
$\tetaot = 1$--$10^\circ$ and $|\Delta m^2_{23}|$ in the range allowed 
by SuperK, a misidentification of the sign of $\Delta m^2_{23}$ 
can be excluded at 99\% CL at 3500 km and 7332 km, but not at the shortest 
distance, 732 km. This conclusion agrees with the analysis of ref. 
\cite{bgrw}, which did not include the energy dependence information. 
We have further studied how the matter parameter $A=\sqrt{2}\, G_F\, n_e$ 
and the angle $\tetaot$ can be measured simultaneously. Fig.~\ref{fig:sma8be} 
shows the result of a $\chi^2$ fit as described in \cite{golden}.  
Statistical errors as well as backgrounds and detection efficiencies 
have been included. At 732 km there is no sensitivity to the 
matter term, as expected. However, already at 3500 km, $A$ can be measured 
with a 10\% precision. At the largest baseline, the precision in $A$ improves 
although at the expense of loosing precision in $\tetaot$ due to the loss 
in statistics. 
%
%

\begin{figure}
\begin{center}
\epsfig{file=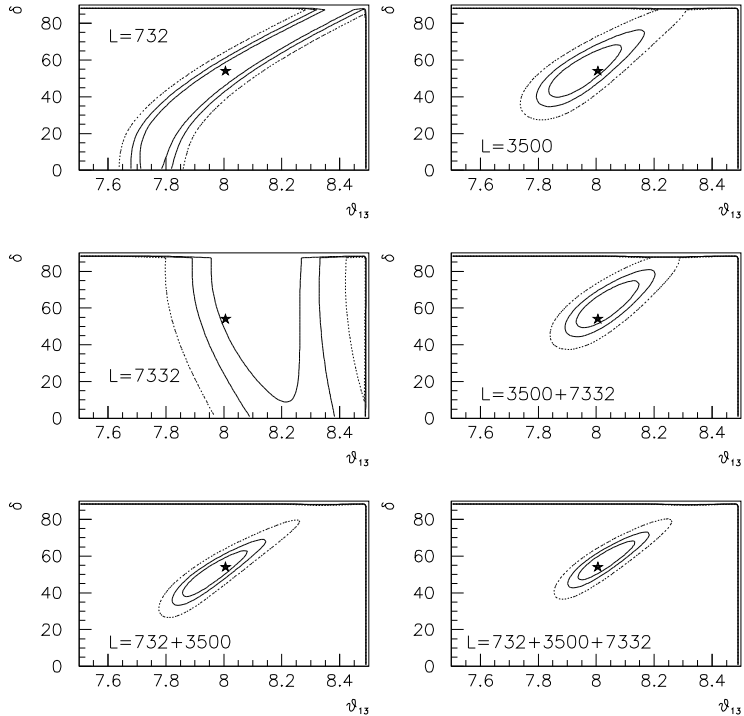,width=11cm} 
\end{center}
\caption{\it 68.5, 90, 99 \% CL contours resulting from a $\chi^2$ fit of  
$\tetaot$ and $\delta$. The parameters used to generate the ``data'' are 
depicted by a star and the baseline(s) which is used for the fit indicated 
in each plot. Statistical errors, backgrounds and efficiencies 
are included.}
\label{fig:54}
\end{figure}
%

\begin{figure}
\begin{center}
\epsfig{file=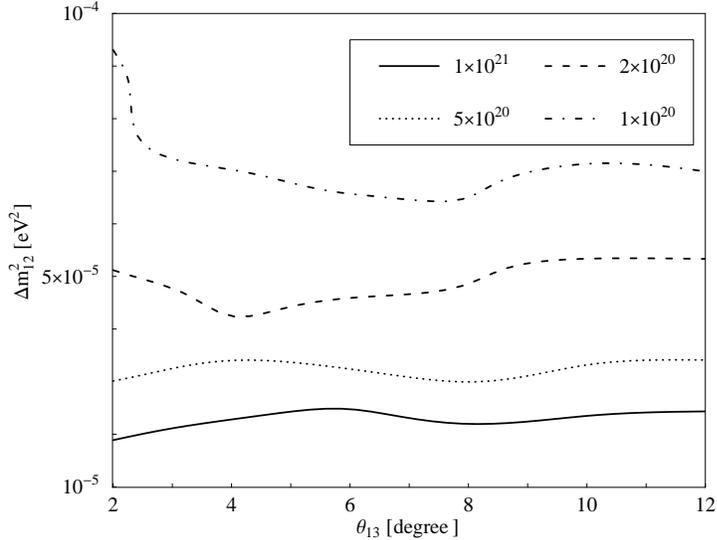,width=10cm}
\end{center} 
\caption{\it Lower limit in $\Delta m_{12}^2$ at which a maximal CP phase 
($90^\circ$) can be distinguished from a vanishing phase at 99\% CL, as a 
function of $\tetaot$ at $L = $ 3500km and for different numbers of 
useful muons per year. Background errors and efficiencies are included.}
\label{fig:limdm12}
\end{figure}

{\bf LMA-MSW.} 
Fixed values of the atmospheric parameters\footnote{A precision of $1\%$ 
in these parameters is achievable through muon disappearance measurements 
at the neutrino factory \cite{bgrw}. This level of uncertainty is not 
expected to affect significantly the results presented in this section.} 
are used in 
this section, $\Delta m_{23}^2 = 2.8\times 10^{-3}$ eV$^2$ and maximal 
mixing, $\tetatt=45^\circ$. Let us start discussing the measurement of 
the CP phase $\delta$ versus $\tetaot$. We have studied numerically how to 
disentangle them in the range $1$--$10^\circ$ and $0 \le \delta \le 180^\circ$. 
Consider first the upper solar mass range allowed by the LMA-MSW solution: 
$\Delta m_{12}^2=10^{-4}$ eV$^2$. Fig.~\ref{fig:54} shows the confidence 
level contours for a simultaneous fit of $\tetaot$ and $\delta$, for ``data'' 
corresponding to $\tetaot=8^\circ$, $\delta=54^\circ$, including in the 
analysis statistical errors as well as backgrounds and detection efficiencies. 
The correlation between $\delta$ and $\tetaot$ is very large. 
The phase $\delta$ is not measurable and this indetermination 
induces a rather large error on the angle $\tetaot$. However, at the 
intermediate baseline of 3500 km the two parameters can be disentangled 
and measured. At the largest baseline, the sensitivity to $\delta$ is lost 
and the precision in $\tetaot$ becomes worse due to the smaller statistics. 
The combination of the results for 3500 km with that for any one of the 
other distances improves the fit, but not in a dramatic way. Just one 
detector located at $O$ (3000 km) may be sufficient: a precision of few 
tenths of degree is attained for $\tetaot$ and of a few tens of degrees 
for $\delta$.
%
%
The sensitivity to CP-violation decreases linearly with $\Delta m_{12}^2$.
At the central value allowed by the LMA-MSW solution, $\Delta m_{12}^2 = 
5\times 10^{-5}$ eV$^2$, CP-violation can still be discovered, 
while for $\Delta m_{12}^2 = 1 \times 10^{-5}$ eV$^2$, the sensitivity to 
CP-violation is lost with the experimental set-up used. 
We have quantified what is the minimum value of $\Delta m^2_{12}$ for which 
a maximal CP-odd phase, $\delta = 90^\circ$, can be distinguished at 99\% CL 
from $\delta = 0^\circ$. The result is shown in Fig.~\ref{fig:limdm12}: 
$\Delta m^2_{12} > 2 \times 10^{-5}$ eV$^2$, with very small dependence 
on $\tetaot$, in the range considered. 

%
%
%

One word of caution is pertinent: up to now we assumed $|\Delta m^2_{12}|$ 
and $\sin 2 \theta_{12}$ known by the time the neutrino factory will be 
operational. Otherwise, the correlation of these parameters with $\tetaot$ 
would be even more problematic than that between $\delta$ and $\tetaot$. 
The error induced on $\tetaot$ by the present uncertainty  
in $|\Delta m^2_{12}|$ is much larger than that stemming from the 
uncertainty on $\delta$. Fortunately, LBL reactor experiments will measure 
$|\Delta m^2_{12}|$ and $\sin 2 \theta_{12}$, if in the LMA-MSW range. 
Even if the error in these measurements is as large as 50\%, the problem 
would be much less serious. We have checked that such uncertainty does
not affect our results concerning the sensitivity to $\delta$, and only
induces an error in $\theta_{13}$ of the order 20\%-30\%.


{\bf Conclusions.} 
We have shown that an analysis in neutrino energy bins, combined with a 
comparison of the signals obtained with the two polarities, allows to 
disentangle the unknown parameters, in particular $\tetaot$ and $\delta$, 
at long enough baselines. Once one takes into account realistic backgrounds 
and detection efficiencies, the intermediate baseline of $O$(3000 km) is 
optimal for the physics goals considered. 

Our two parameter fits at 3500 km indicate that
the angle $\tetaot$ can be measured with a precision of tenths of degrees, 
down to values of $\tetaot=1^\circ$. The asymptotic sensitivity to 
$\sin^2 \tetaot$ can be improved by three orders of magnitude (or more) 
with respect to the present upper bound.

In the LMA-MSW range, CP-violation may be tackled. The phase $\delta$ can 
be determined with a precision of tens of degrees, for the central values 
allowed for $|\Delta m_{12}^2|$. Maximal CP-violation can be disentangled 
from no CP-violation at 99\% CL for values of $|\Delta m_{12}^2| > 
2 \times 10^{-5}$ eV$^2$. 

Finally, a model independent confirmation of the 
MSW effect will be feasible, and the matter parameter $A$ measured within a 
10\% precision (or better if combined with the longest baseline: $7332$ km). 
The sign of $\Delta m_{23}^2$, can be determined at 99\% CL, for $\tetaot$ 
within the range $\tetaot=1$--$10^\circ$ and $|\Delta m_{23}^2|$ in the 
range allowed by SuperK.
%
%
%

%
%

\begin{thebibliography}{9}
%
\bibitem{Superka} 
Y.~Fukuda {\em et al.}, Phys. Lett. {\bf B 433} (1998) 9, ibid. 
{\bf B 436} (1998) 33 and ibid. {\bf B 467} (1999) 185; 
Phys. Rev. Lett. {\bf 82} (1999) 2644; 
T.~Kajita, Nucl. Phys. {\bf B} (Proc. Suppl.) {\bf 77} (1999) 123.
%
\bibitem{otheratm}
S.~Hatakeyama {\em et al.}, Phys. Rev. Lett. {\bf 81} (1998) 2016;
E.~Peterson, Nucl. Phys. {\bf B} (Proc. Suppl.) {\bf 77} (1999) 111; 
W.~W.~Allison, Phys. Lett. {\bf B 449} (1999) 137; 
F.~Ronga {\em et al.}, Nucl. Phys. {\bf B} (Proc. Suppl.) {\bf 77} (1999) 117.
%
\bibitem{sol} 
R.~Davis {\em et al.}, Phys. Rev. Lett. {\bf 20} (1968) 1205; 
B.~T.~Cleveland {\em et al.}, Astrophys. J. {\bf 496} (1998) 505; 
K.~Lande {\em et al.}, Nucl. Phys. (Proc. Suppl.) {\bf 77} (1999) 13; 
T.~Kirsten, Nucl. Phys. (Proc. Suppl.) {\bf 77} (1999) 26; 
Y.~Fukuda {\em et al.}, Phys. Rev. Lett. {\bf 77} (1996) 1683; 
Dzh.~N.~Abdurashistov {\em et al.}, Nucl. Phys. (Proc. Suppl.) {\bf 77} 
(1999) 20; 
Y.~Fukuda {\em et al.}, Phys. Rev. Lett. {\bf 82} (1999) 1810; 
M.~Smy {\em et al.}, hep-ex/9903034.
%
%
%
\bibitem{newSK}
Y. Suzuki and H. Sobel talks at Neutrino2000, 16-21 June, Subdury, CANADA.
%
\bibitem{us99} 
A.~Donini et {\em al.}, 
Nucl. Phys. {\bf B574} (2000) 23, 
hep-ph/9910516 
and hep-ph/0007283. 
%
\bibitem{geer} 
S.~Geer, Phys. Rev. {\bf D 57} (1998) 6989.
%
\bibitem{firstmachine} 
C.~M.~Ankenbrandt {\em et al.} (Muon Collider Collaboration), 
Phys. Rev. ST Accel. Beams {\bf 2}, (1999) 081001; 
B.~Autin {\em et al.}, CERN-SPSC/98-30, SPSC/M 617 (October 1998); 
S.~Geer, C.~Johnstone and D.~Neuffer, FERMILAB-PUB-99-121.
%
\bibitem{dgh}
A.~de~R\'ujula, M.~B.~Gavela and P.~Hern\'andez, Nucl. Phys. {\bf B 547} 
(1999) 21. 
%
\bibitem{golden} For a complete review of the subject presented in this 
talk and for a complete set of refereces see: 
A.~Cervera et {\em al.}, 
Nucl.\ Phys.\  {\bf B579}, 17 (2000).
%
\bibitem{others}
See also K.~Dick et {\em al.}, Nucl. Phys. {\bf B 562} (1999) 29; 
M.~Campanelli et {\em al.}, hep-ph/9905240 and hep-ph/0005007; 
V.~Barger et {\em al.}, Phys. Rev. {\bf D 61} (2000) 053004;
M.~Freund {\em al.}, Nucl.Phys. {\bf B578} (2000) 27.
%
\bibitem{cdg}
A.~Cervera, F.~Dydak and J.~G\'omez-Cadenas, Nufact'99 and Nufact'00 
Whorkshops.
%
\bibitem{bgrw} 
V.~Barger et {\em al.}, Phys. Rev. {\bf D62}, 013004 (2000);
%
\end{thebibliography}
\end{document}